\documentclass[aps,pre,twocolumn,amsmath,amsfonts,showpacs]{revtex4-1}
\newcommand{\bec}[1]{\mbox{\boldmath $ #1$}}
\usepackage{graphicx}
\begin{document}
\title{Transition phenomena in unstably stratified turbulent flows}
\author{M. Bukai}
\email{mbukai@gmail.com}
\author{A. Eidelman}
\email{eidel@bgu.ac.il}
\author{T. Elperin}
\email{elperin@bgu.ac.il}
\homepage{http://www.bgu.ac.il/me/staff/tov}
\author{N. Kleeorin}
\email{nat@bgu.ac.il}
\author{I. Rogachevskii}
\email{gary@bgu.ac.il}
\homepage{http://www.bgu.ac.il/~gary}
\author{I. Sapir-Katiraie}
\email{katiraie@gmail.com}
\affiliation{The Pearlstone Center for Aeronautical Engineering
Studies, Department of Mechanical Engineering,
Ben-Gurion University of the Negev, P.O.Box 653, Beer-Sheva 84105,  Israel}
\date{\today}
\begin{abstract}
We study experimentally and theoretically transition phenomena caused by the external forcing from Rayleigh-B\'{e}nard convection with the large-scale circulation (LSC) to the limiting regime of unstably stratified turbulent flow without LSC whereby the temperature field behaves like a passive scalar. In the experiments we use the Rayleigh-B\'{e}nard apparatus with an additional source of turbulence produced by two oscillating grids located nearby the side walls of the chamber. When the frequency of the grid oscillations is larger than $2$ Hz, the large-scale circulation (LSC) in turbulent convection is destroyed, and the destruction of the LSC is accompanied by a strong change of the mean temperature distribution. However, in all regimes of the unstably stratified turbulent flow the ratio $\big[(\ell_x \nabla_x T)^2 + (\ell_y \nabla_y T)^2 + (\ell_z \nabla_z T)^2\big] / \langle \theta^2 \rangle$ varies slightly (even in the range of parameters whereby the behaviour of the temperature field is different from that of the passive scalar). Here $\ell_i$ are the integral scales of turbulence along $x, y, z$ directions, $T$ and $\theta$ are the mean and fluctuating parts of the fluid temperature. At all frequencies of the grid oscillations we have detected the long-term nonlinear oscillations of the mean temperature. The theoretical predictions based on the budget equations for turbulent kinetic energy, turbulent temperature fluctuations and turbulent heat flux, are in agreement with the experimental results.
\end{abstract}


\maketitle

\section{Introduction}

Various aspects of formation of large-scale circulations (LSC) and other types of coherent structures in a turbulent convection at large Rayleigh numbers have been discussed in a number of studies for the atmospheric flows \cite{EB93,AZ96,Z91,EKRZ06}, laboratory experiments in the Rayleigh-B\'{e}nard apparatus
\cite{KH81,SWL89,TBL93,CCL96,NSS01,SB02,NS03,QT01,XLZ02,BKT03,XSZ03,LLT05,XL04,SQ04,FA04,BNA05,VTH06,EEKR06,RPT06,BA06,PINT07,XX07,XX09,PRT09,HT09,BOD09}
and direct numerical simulations \cite{KER01,HTB03,PHP04,RJH05}.
A mean-field theory \cite{EKRZ02,EKRZ06,EGKR06} has been proposed for the explanation of formation of coherent structures in turbulent convection. It is known from laboratory experiments and numerical simulations \cite{TBL93,BKT03,XSZ03,VTH06} that the coherent structures in  the Rayleigh-B\'{e}nard turbulent convection are not driven by the turbulent Reynolds stresses, associated with the tilting plumes at the upper and the lower horizontal walls. If the mean flow is established, the temperature of the fluid is larger at one side of the LSC and smaller at the other side of the LSC, and the mean flow is driven by the mean buoyancy force at the side walls of the chamber.

The effect of coherent structures (mean wind) on global properties of turbulent convection is a subject of numerous discussions (see, e.g., reviews \cite{SI94,K01,AGL09}). It has been recently found in laboratory experiments in \cite{BEKR09}, that in turbulent convection with the mean wind the bulk production of turbulence is due to the large-scale shear caused by the mean wind rather than by buoyancy.

Numerous experiments demonstrate that the mean temperature distribution in turbulent convection is strongly inhomogeneous and anisotropic due to the mean wind (see, e.g., \cite{TBL93,XSZ03,SQ04,VTH06,BEKR09}), and the mean temperature gradient in the horizontal direction inside the LSC can be significantly larger than in the vertical direction, i.e., the vertical turbulent heat flux inside the LSC is very small. The dependencies of the dissipation rate (normalized by the molecular temperature diffusion) of the temperature fluctuations $\langle \theta^2 \rangle$ as a function of the Rayleigh number and different statistical characteristics of this normalized dissipation rate have been studied experimentally for turbulent Rayleigh-B\'{e}nard convection in \cite{HT09}.

On the other hand, in a forced turbulent convection without a mean wind (i.e., in unstably stratified turbulent flows) the temperature field behaves like a passive scalar (see, e.g., \cite{SG90,SI94,AS99,LX10}). It is not clear why in turbulent convection with LSC the vertical gradient of the mean temperature field is small inside LSC, and the vertical temperature gradient can even change the direction, i.e., it can be directed upwards. A detailed study of a transition from one regime with the mean wind to another regime without the mean wind due to additional forcing in unstably stratified turbulent flow, can elucidate the effects of the mean wind on turbulent convection.

Different effects in turbulent convection with a forcing have been studied previously (see, e.g., \cite{SG90,JX08,GI06,GI09}). In particular, an experimental study of a turbulent convection with forcing in the form of periodical pulses has been performed in \cite{JX08}. In these experiments, in spite of the forcing, a large-scale circulatory flow was not destroyed, and the heat transfer in this system was enhanced in comparison with constant and sinusoidally modulated energy inputs. It was found in \cite{JX08} that the enhancement of the heat transfer depends on the synchronization of the kicking period of energy input with the intrinsic time scale of the turbulent flow. In other type of experiments \cite{GI06,GI09}
a relation between convective heat flux and temperature gradient was determined in a vertical channel filled with water, the average vertical mass flux being zero. It was shown in these experiments that when the LSC is suppressed by physically blocking its path, there exists a strong vertical temperature gradient in the bulk of the Rayleigh-B\'{e}nard apparatus and the temperature remains active, rather than passive field.

The goal of this paper is to study experimentally and theoretically transition phenomena due to the external forcing from Rayleigh-B\'{e}nard convection with LSC to the limiting regime of unstably stratified turbulent flow without LSC, whereby the temperature field behaves like a passive scalar. In the experimental study we use  the Rayleigh-B\'{e}nard apparatus with an additional source of turbulence produced by the two oscillating grids located nearby the side walls of the chamber. Additional forcing for turbulence allows us to observe evolution of the mean temperature and velocity fields during the transition from turbulent convection with the LSC (for very small frequencies of the grid oscillations) to the limiting regime of unstably stratified flow without LSC (for very high frequencies of the grid oscillations). In the experiments we use Particle Image Velocimetry to determine the turbulent and mean velocity fields, and a specially designed temperature probe with sensitive thermocouples is employed
to measure the temperature field.

Note that this experimental set-up with the turbulent convection and the external forcing has been previously used in studies of turbulent transport of particles (see, e.g., \cite{BEE04,EE04,EEKR06C,EEKR06A}). Comprehensive investigation  of turbulence structures, mean velocity and temperature distributions, velocity and temperature fluctuations can elucidate a complicated physics related to particle clustering and formation of large-scale inhomogeneities in particle spatial distributions in unstably stratified turbulent flows.
Notably, turbulent convection with the external forcing is used in chemical industrial devices to improve mixing (see, e.g., \cite{BB99,CS98,BR99}). Another example of turbulent convection with forcing is atmospheric wind superimposed on turbulent convective boundary layer.

The paper is organized as follows. Section II
describes the experimental set-up, instrumentation and the results of laboratory study of the transition phenomena in unstably stratified turbulent flow. The theoretical analysis and comparison with the experimental results are performed in Section III. Finally, conclusions are drawn in Section IV. In Appendix we present the detailed derivations of the theoretical model for a passive scalar that used for the explanation of the observed effects.

\section{Experimental study}

Let us start with description of the experimental set-up and the results of laboratory study of the transition phenomena in unstably stratified turbulent flow.

\subsection{Experimental set-up and instrumentation}

\begin{figure}
\vspace*{1mm}
\centering
\includegraphics[width=8cm]{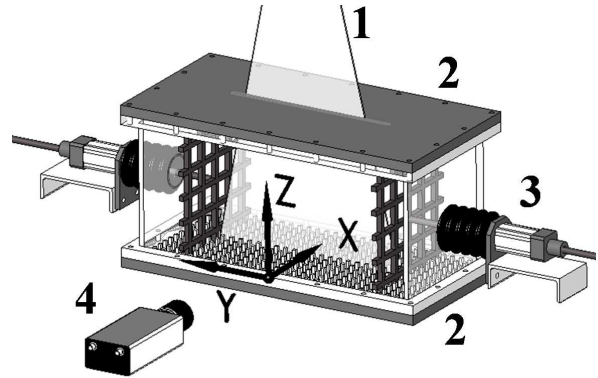}
\caption{\label{Fig1} Experimental set-up: (1) -laser light sheet; (2) - heat exchangers; (3) - grid driver; (4) - digital CCD camera.}
\end{figure}

The experiments on transition phenomena in unstably stratified turbulence have been conducted in rectangular chamber with dimensions $26 \times 58 \times 26$ cm$^3$ in air flow with the Prandtl number Pr $=0.71$.  The side walls of the chambers are made of transparent Perspex with the thickness of $10$ mm.
In the experiments we use the Rayleigh-B\'{e}nard apparatus with an additional source of turbulence produced by two oscillating grids. Pairs of vertically oriented grids with bars arranged in a square array (with a mesh size 5 cm) are attached to the right and left horizontal rods (see Fig.~\ref{Fig1}). The grids are positioned at a distance of two grid meshes from the chamber walls and are parallel to the side walls. Both grids are operated at the same amplitude of $61$ mm, at a random phase and at the same frequency which is varied in the range from $0.5$ Hz to $16.5$ Hz. Here we use the following system of coordinates: $z$ is the vertical axis, the $y$-axis is perpendicular to the grids and the $xz$-plane is parallel to the grids.

A vertical mean temperature gradient in the turbulent air flow was
formed by attaching two aluminium heat exchangers to the bottom and
top walls of the test section (a heated bottom and a cooled top wall
of the chamber). In order to improve heat transfer in the boundary layers at the bottom and top walls we used heat exchangers with rectangular fins $3 \times 3 \times 15$ mm$^3$ (see Fig.~\ref{Fig1}) which allowed us to form a mean temperature gradient up to 115 K/m at a mean temperature of about 308 K when the frequency of the grid oscillations $f > 10$ Hz. The thickness of the aluminium plates with the fins is 2.5 cm. The top plate is a bottom wall of the tank with cooling water.
Temperature of water circulating through the tank and the chiller is
kept constant within $0.1$ K. Cold water is pumped into the cooling system through two inlets and flows out through two outlets located at the side wall of the cooling system. The bottom plate is attached to the electrical heater that provides constant and uniform heating. The
voltage of a stable power supply applied to the heater varies up to 200 V. The power of the heater varies up to 300 W.

The temperature field was measured with a temperature probe equipped
with twelve E-thermocouples (with the diameter of 0.13 mm and the
sensitivity of $\approx 65 \, \mu$V/K) attached to a vertical rod with a diameter 4 mm. The spacing between thermal couples along the rod was
22 mm. Each thermocouple was inserted into a 1 mm diameter and 45 mm
long case. A tip of a thermocouple protruded at the length of 15 mm
out of the case. The temperature was measured for 12 rod
positions with 23 mm intervals in the horizontal direction, i.e., at
144 locations in a flow. The exact position of each thermocouple was
measured using images captured with the optical system employed in
PIV measurements. A sequence of 600 temperature readings for every
thermocouple at every rod position was recorded and processed using
the developed software based on LabView 7.0. Temperature maps were plotted using Matlab~7.

In order to decrease mean flow generated by oscillating grids and to increase the range of homogeneous and isotropic turbulence we use partitions located in $xz$ plane between grids and side walls of the chamber (of the width 85 mm). The aspect ratio of the chamber $A =H_y/H_z = 1.52$, where $H_y$ is the size of the chamber along $y$-axis between partitions and $H_z$ is the hight of the chamber $z$-axis, respectively.

The velocity fields were measured using a Stereoscopic Particle Image Velocimetry (PIV), see \cite{AD91,RWK98,W00}. In the experiments we used LaVision Flow Master III system. A double-pulsed light sheet was provided by a Nd-YAG laser (Continuum Surelite $ 2 \times 170$ mJ). The light sheet optics includes spherical and cylindrical Galilei telescopes with tuneable divergence and adjustable focus length. We used two
progressive-scan 12 bit digital CCD cameras (with pixel size $6.7 \,
\mu$m $\times \, 6.7 \, \mu$m and $1280 \times 1024$ pixels) with a
dual-frame-technique for cross-correlation processing of captured
images. A programmable Timing Unit (PC interface card) generated
sequences of pulses to control the laser, camera and data
acquisition rate. The software package LaVision DaVis 7 was applied
to control all hardware components and for 32 bit image acquisition
and visualization. This software package comprises PIV software for
calculating the flow fields using cross-correlation analysis.

In order to obtain velocity maps in the central region of the flow in the cross-section parallel to the grids and perpendicular to a front view plane, we used one camera with a single-axis Scheimpflug adapter. The angle between the optical axis of the camera and the front view plane as well as the angle between the optical axis and the probed cross-section was approximately 45 degrees. The perspective distortion was compensated using Stereoscopic PIV system calibration kit whereby the correction was calculated for a recorded image of a calibration plate. The corrections were introduced in the recorded in the probed cross-section images before their processing using a cross-correlation technique for determining velocity fields.

An incense smoke with sub-micron particles ($\rho_p / \rho \sim 10^3)$, as a tracer was used for the PIV measurements. Smoke was produced by high temperature sublimation of solid incense grains. Analysis of smoke particles using a microscope (Nikon, Epiphot with an amplification of 560) and a PM-300 portable laser particulate analyzer showed that these particles have an approximately spherical shape and that their mean diameter is of the order of $0.7 \mu$m. The probability density
function of the particle size measured with the PM300 particulate
analyzer was independent of the location in the flow for incense
particle size of $0.5-1 \, \mu $m. The maximum tracer particle displacement in the experiment was of the order of $1/4$ of the interrogation window. The average displacement of tracer particles was of the order of $2.5$ pixels. The average accuracy of the velocity measurements was of the order of $4 \%$ for the accuracy of the correlation peak detection in the interrogation window of the order of $0.1$ pixel (see, e.g., \cite{AD91,RWK98,W00}).

We determined the mean and the r.m.s.~velocities, two-point
correlation functions and an integral scale of turbulence from the
measured velocity fields. Series of 520 pairs of images acquired with a frequency of 1 Hz, were stored for calculating velocity maps and for ensemble and spatial averaging of turbulence characteristics. The center of the measurement region in $yz$-plane coincides with the center of the chamber. We measured velocity in a flow domain
$256 \times 256$ mm$^2$ with a spatial resolution of $1024 \times 1024$ pixels. This corresponds to a spatial resolution 250 $\mu$m / pixel.
The velocity field in the probed region was analyzed with interrogation windows of $32 \times 32$ or $16 \times 16$ pixels. In every
interrogation window a velocity vector was determined from which
velocity maps comprising $32 \times 32$ or $64 \times 64$ vectors
were constructed. The mean and r.m.s. velocities for every point of
a velocity map were calculated by averaging over 520
independent maps, and then they were averaged over the central flow region.

The two-point correlation functions of the velocity field were
determined for every point of the central part of the velocity map
(with $16 \times 16$ vectors) by averaging over 520 independent
velocity maps, which yields 16 correlation functions in $y$ and $z$ directions. Then the two-point correlation function was obtained by averaging over the ensemble of these correlation functions. An integral scale of turbulence, $\ell$, was determined from the two-point correlation functions of the velocity field. In the experiments we evaluated the variability between the first and the last 20 velocity maps of the series of the measured velocity field. Since very small variability was found, these tests showed that 520 velocity maps contain enough data to obtain reliable statistical estimates. We found also that the measured mean velocity field is stationary.

The characteristic turbulence time in the experiments $\tau_z = 0.28 - 0.62$ seconds, while the characteristic time period for the large-scale circulatory flow $(\sim 10$ s) is by 1.5 orders of magnitude larger than $\tau_z$. These two characteristic times are much smaller than the time during which the velocity fields are measured $(\sim 600$ s). The size of the probed region did not affect our results. The temperature difference  between the top and bottom plates, $\Delta T$, in all experiments was 50 K (i.e., the global Rayleigh number based on molecular transport coefficients, was ${\rm Ra} = 0.73 \times 10^8$). Similar experimental set-up and data processing procedure were used previously in the experimental study of different aspects of turbulent convection \cite{EEKR06,BEKR09} and in \cite{BEE04,EE04,EEKR06C,EEKR06A} for investigating a phenomenon of turbulent thermal diffusion \cite{EKR96,EKRS00}.

\subsection{Experimental results}

\begin{figure}
\vspace*{1mm}
\centering
\includegraphics[width=7.5cm]{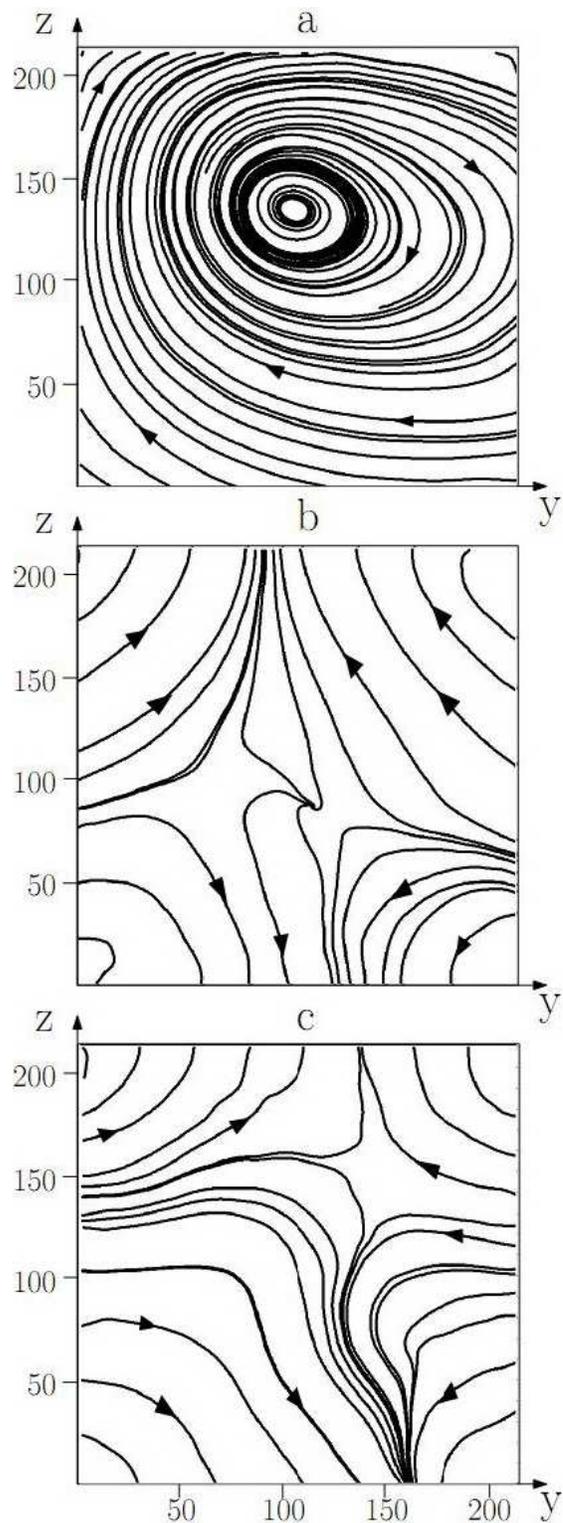}
\caption{\label{Fig2} Mean flow patterns obtained in the experiments with unforced turbulent convection (upper panel) and with forced turbulent convection at the frequencies $f=2.2$ Hz (middle panel) and $f=16.5$ Hz (lower panel) of the grid oscillations for the unstably stratified turbulent flow at the temperature difference between the top and bottom walls $\Delta T = 50$ K. Coordinates $y$ and $z$ are measured in mm.}
\end{figure}

Let us start the analysis of the obtained experimental results from the dynamics of the mean velocity field. In Fig.~\ref{Fig2} we show the mean flow patterns obtained in the experiments with unforced turbulent convection (Fig.~\ref{Fig2}a) and with forced turbulent convection at different frequencies of the grid oscillations (Figs.~\ref{Fig2}b and~\ref{Fig2}c). The solid lines in Fig.~\ref{Fig2} are stream lines in the cental $yz$-plane for $x=13$ cm. The center of Figs.~\ref{Fig2}a,  \ref{Fig2}b, \ref{Fig2}c, which has coordinates $y=10$ cm and $z=10$ cm, coincides with the center of the chamber. When the frequency of the grid oscillations is larger than $2$ Hz, the large-scale circulation (LSC) in turbulent convection in the $yz$-plane is not observed (see Figs.~\ref{Fig2}b and~\ref{Fig2}c). On the other hand, a complicated mean flow of the spiral form in the $xz$-plane with a non-zero mean vorticity in the $y$ direction exists when the frequencies of the grid oscillations $2<f<7$ Hz.

\begin{figure}
\vspace*{1mm}
\centering
\includegraphics[width=9cm]{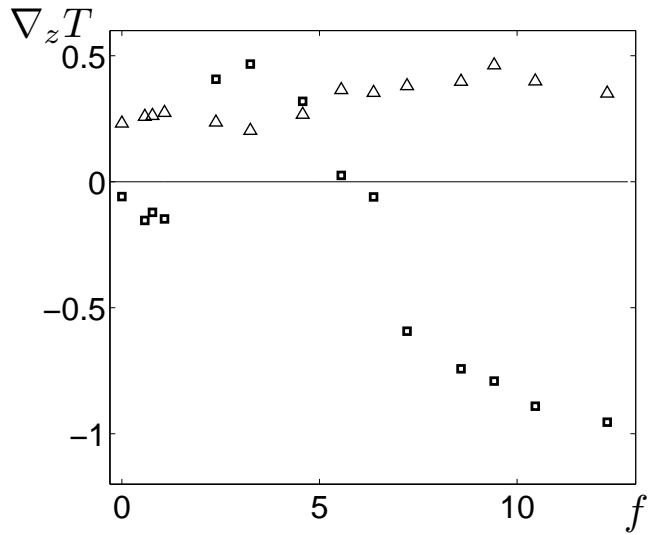}
\caption{\label{Fig3} Vertical gradient of the mean temperature $\nabla_z T$ (squares) and the amplitude of the nonlinear long-term oscillations of this gradient (triangles) versus the frequency $f$ of the grid oscillations for the unstably stratified turbulent flow. The mean temperature gradient is measured in K cm$^{-1}$ and the frequency $f$ is measured in Hz.}
\end{figure}

\begin{figure}
\vspace*{1mm}
\centering
\includegraphics[width=9cm]{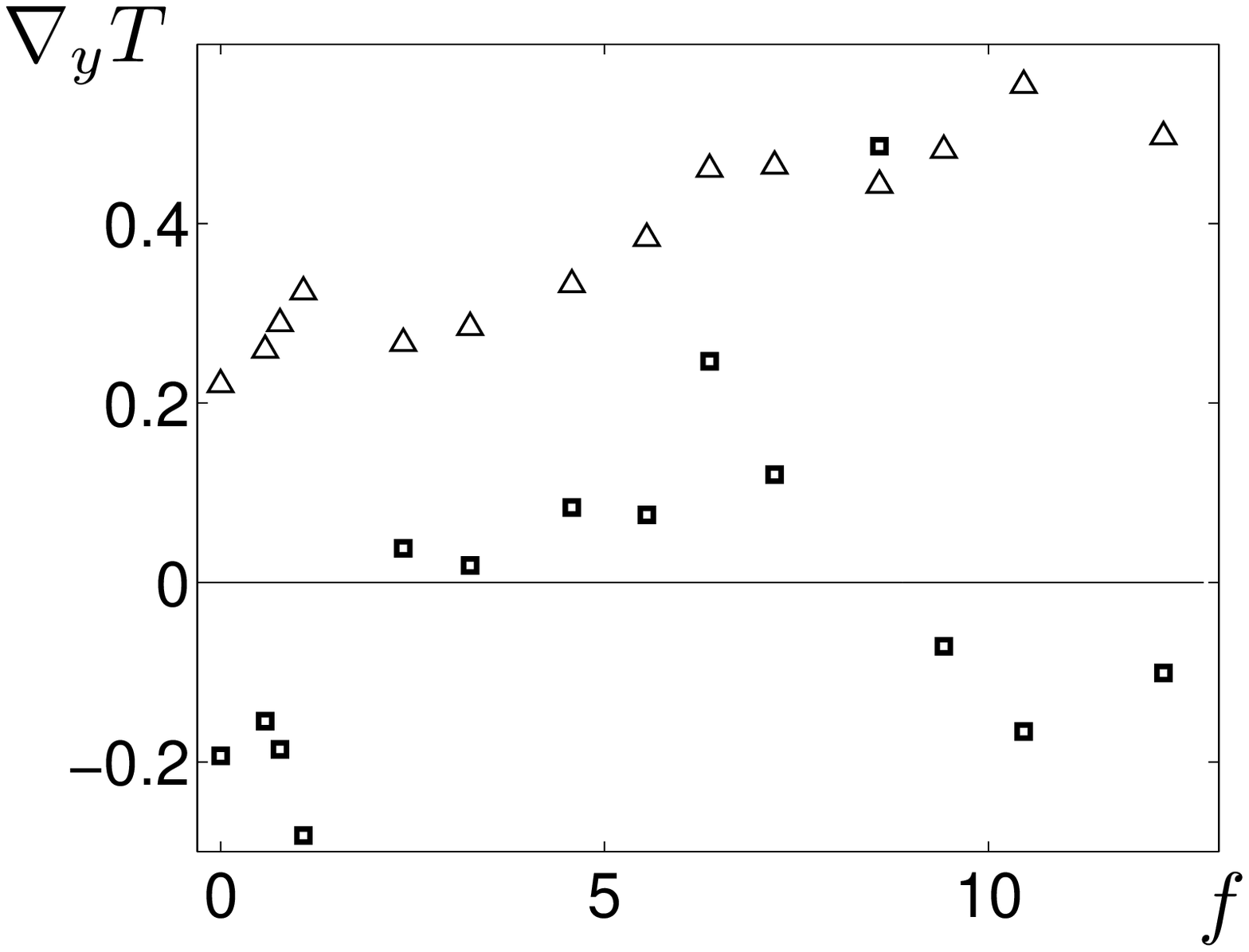}
\caption{\label{Fig4} Horizontal gradient of the mean temperature $\nabla_y T$ (squares) and the amplitude of the nonlinear long-term oscillations of this gradient (triangles) versus the frequency $f$ of the grid oscillations for the unstably stratified turbulent flow. The mean temperature gradient is measured in K cm$^{-1}$ and the frequency $f$ is measured in Hz.}
\end{figure}

\begin{figure}
\vspace*{1mm}
\centering
\includegraphics[width=9cm]{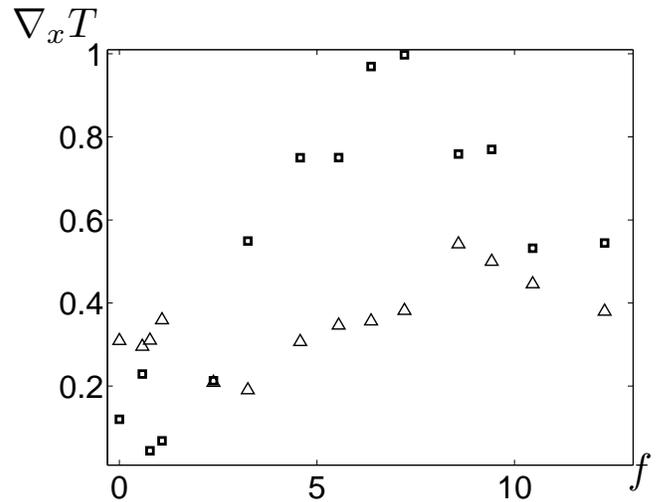}
\caption{\label{Fig5} Horizontal gradient of the mean temperature $\nabla_x T$ (squares) and the amplitude of the nonlinear long-term oscillations of this gradient (triangles) versus the frequency $f$ of the grid oscillations for the unstably stratified turbulent flow. The mean temperature gradient is measured in K cm$^{-1}$ and the frequency $f$ is measured in Hz.}
\end{figure}

The destruction of the LSC is accompanied by a strong change of the mean temperature distribution. The distribution of the measured mean temperature field depends strongly on the frequency $f$ of the grid oscillations (see Figs.~\ref{Fig3}-\ref{Fig5}). In particular, for the very low frequency $f$ the thermal structure inside the LSC in turbulent convection is inhomogeneous and anisotropic. The hot thermal plumes concentrate at one side of LSC, and cold plumes accumulate at the opposite side of LSC \cite{SQ04,BEKR09}. In the central part of the flow the vertical mean temperature gradient $\bec{\nabla_z} T$ changes its sign depending on the frequency of the grid oscillations (see Fig.~\ref{Fig3}).

At all frequencies of the grid oscillations we have observed the long-term nonlinear oscillations of the mean temperature with the period that is of the order of 10 s. Note that the temperature was measured at 144 locations in a flow (the spacing between thermal couples was 22 mm). We perform the sliding averaging of the instantaneous temperature field over time of 3 s (the characteristic turbulence time in the experiments was 0.3 - 0.6 s) in order to determine the mean temperature field $T$ at 144 locations. Then we determine the long-term variations of the vertical mean temperature gradient $\delta (\nabla_i T) = \nabla_i T - \overline{\nabla_i T}$ due to the nonlinear oscillations of the mean temperature, where $i = x, y, z$. The bar here denotes additional time averaging. Note that the separation distance of 22 mm between thermal couples is sufficient to measure the gradients of the mean temperature. Indeed, the integral scale of turbulence is about 2 cm and the characteristic length scale of the mean temperature field is much larger than the integral scale of turbulence. Note also that for large frequencies of the grid oscillations the role of the thermal boundary layer in turbulent convection diminishes. It must be emphasized that in all our experiments rectangular fins were attached at the bottom and top walls of the chamber in order to improve heat transfer in the boundary layers.

\begin{figure}
\vspace*{1mm}
\centering
\includegraphics[width=9cm]{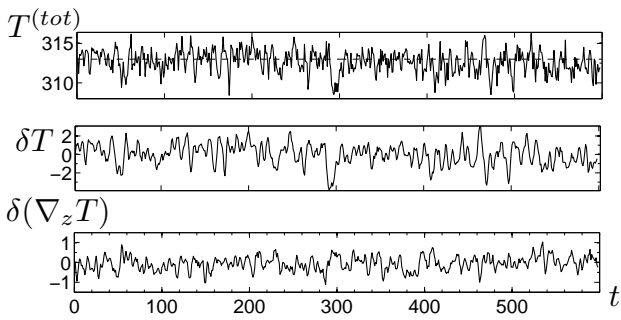}
\caption{\label{Fig6} Time dependencies of the instantaneous temperature $T^{tot}= T+\theta$, the variations of mean temperature $\delta T = T - \overline{T}$ and the variations of the vertical mean temperature gradient $\delta (\nabla_z T) = \nabla_z T - \overline{\nabla_z T}$ due to the long-term nonlinear oscillations of the mean temperature (with the period $\sim 10$ s). The bar denotes additional time averaging. These time dependencies are measured in the center of the chamber at the frequency $f=10$ Hz of the grid oscillations for the unstably stratified turbulent flow. These temperature characteristics are measured in K.}
\end{figure}

In Fig.~\ref{Fig6} we show time dependencies of the instantaneous temperature $T^{tot}=T + \theta$, the long-term variations of mean temperature $\delta T = T - \overline{T}$ and the long-term variations of the vertical mean temperature gradient $\delta (\nabla_z T) = \nabla_z T - \overline{\nabla_z T}$ due to the nonlinear oscillations of the mean temperature. In Figs.~\ref{Fig3}-\ref{Fig5} we also show by triangles the amplitude, $\{\overline{[\delta (\nabla_i T)]^2}\}^{1/2}$, of the nonlinear long-term oscillations of the gradients of the mean temperature, where $i=x, y, z$. The dependencies shown in Figs.~\ref{Fig3}-\ref{Fig6} are obtained at the central region of the chamber with the size of $10 \times 10 \times 10$ cm$^3$.

\begin{figure}
\vspace*{1mm}
\centering
\includegraphics[width=8cm]{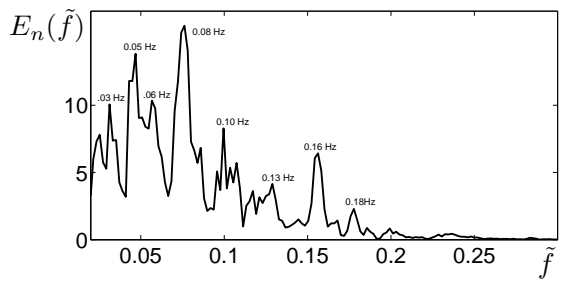}
\caption{\label{Fig7} The normalized spectrum function $E_n(\tilde f) = |(\delta T)_{\tilde f}|^2 / \int |(\delta T)_{\tilde f}|^2 \,d\tilde f$
of the signal $\delta T = T(t) - \overline{T}$, where in the Fourier space $(\delta T)_{\tilde f} = \int \delta T \, \exp[- i  \tilde f \, t] \,d t$ and $\tilde f$ is the frequency of the nonlinear long-term oscillations of the mean temperature.}
\end{figure}

In Fig.~\ref{Fig7} we show the results of a Fourier analysis of the signal $\delta T = T - \overline{T}$. Inspection of Fig.~\ref{Fig7} shows that there are two main maxima in the spectrum with the periods 12.5 s and 20 s. Other smaller maxima in the spectrum are at the frequencies which are multiples of these main frequencies or their sums and differences. These are typical features of nonlinear oscillations. The theory that explains the mechanism of these nonlinear oscillations of the mean temperature field, has not been developed yet. One may hypothesize that there are two possible mechanisms for such oscillations. One mechanism could be related to the large-scale  Tollmien-Schlichting waves in sheared turbulent flows (see \cite{EGKR07}), which can cause the nonlinear oscillations of the mean temperature field. Another mechanism of the nonlinear oscillations could be related to the generation of small-scale kinetic helicity due to large-scale shear flows in the system. The large-scale shear generates large-scale helicity, and since the total helicity is conserved, a non-zero small-scale helicity is produced. Our preliminary analysis shows that the small-scale helicity can cause the nonlinear oscillations of the mean temperature field.

\begin{figure}
\vspace*{1mm}
\centering
\includegraphics[width=9cm]{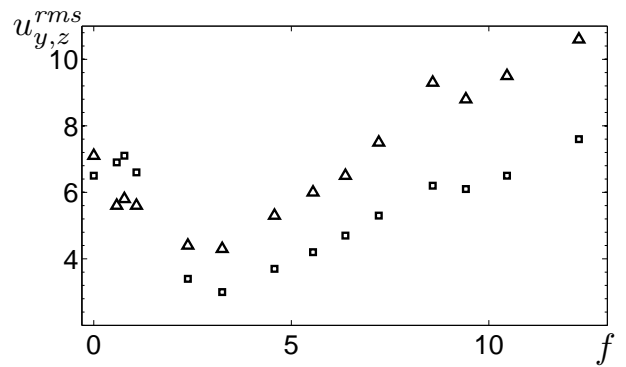}
\caption{\label{Fig8} Components $u_y^{rms} = \sqrt{\langle u_y^2 \rangle}$ (triangles) and $u_z^{rms} = \sqrt{\langle u_z^2 \rangle}$ (squares) of the r.m.s. velocity versus the frequency $f$ of the grid oscillations for the unstably stratified turbulent flow. The turbulent velocity is measured in cm s$^{-1}$ and the frequency $f$ is measured in Hz.}
\end{figure}

\begin{figure}
\vspace*{1mm}
\centering
\includegraphics[width=9cm]{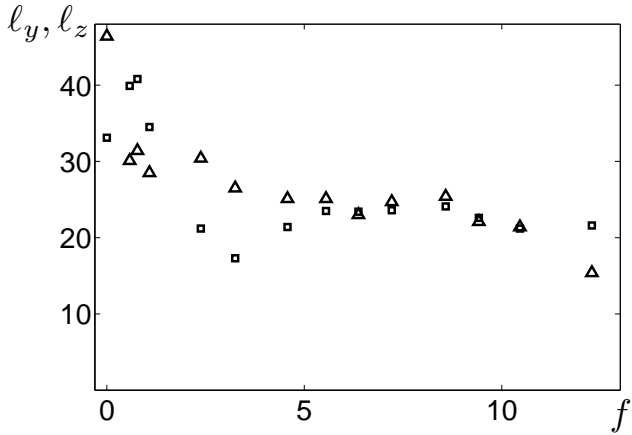}
\caption{\label{Fig9} Horizontal and vertical integral scales of turbulence $\ell_y$ (triangles) and $\ell_z$ (squares) versus the frequency $f$ of the grid oscillations for the unstably stratified turbulent flow. The turbulent length scales are measured in mm and the frequency $f$ is measured in Hz.}
\end{figure}

In order to avoid side effects of the grids we present the experimental results recorded in the central region of the chamber with the size of $10 \times 10 \times 10$ cm$^3$. In Figs.~\ref{Fig8}-\ref{Fig9} we show the frequency dependencies of the following measured turbulence parameters: the r.m.s. velocity fluctuations $\sqrt{\langle u_y^2 \rangle}$ and $\sqrt{\langle u_z^2 \rangle}$, and the corresponding integral scales of turbulence along horizontal $y$ and vertical $z$ directions $(\ell_y$ and $\ell_z)$, where ${\bf u}$ are the fluctuations of the fluid velocity. The mean, r.m.s. velocities and the two-point correlation functions of the velocity field have been calculated by averaging over 520 independent velocity maps, and then they have been averaged over the central flow region. The integral scales of turbulence, $\ell_y$ and $\ell_z$, have been determined from the normalized two-point longitudinal correlation functions of the velocity field, e.g., $F_y(\tilde y)=\langle u_y({\bf r}_0) \, u_y({\bf r}_0 + \tilde y \, {\bf e}_y)\rangle / \langle u_y^2({\bf r}_0)\rangle$ [and similarly for $F_z(\tilde z)$ after replacement in the above formula $y$ by $z$], using the following expression: $\ell_y = \langle \int_0^L F_y (\tilde y) \, d \tilde y \rangle_S$ [and similarly for $\ell_z$ after replacement in the above formula $y$ by $z$], where $L=10$ cm is the linear size of the probed flow region, ${\bf e}_y$ is the unit vector in the $y$ direction, and $\langle ... \rangle_S$ is the additional averaging over the $yz$ plane.

Inspection of Fig.~\ref{Fig9} shows that the turbulent length scales are weakly dependent on the frequency $f$ of the grid oscillations in the unstably stratified turbulent flow for $f > 3$ Hz, while the turbulent velocities and turbulent time scales vary strongly with the frequency $f$. Note that $\tau_{y} \, f$ and $\tau_{z} \, f$ tend to a constant ($\sim 3$) for higher frequencies ($f > 5$ Hz) of the forcing, where $\tau_{y}=\ell_y/\sqrt{\langle u_y^2 \rangle}$ and $\tau_{z}=\ell_z/\sqrt{\langle u_z^2 \rangle}$ are the characteristic turbulent times in horizontal and vertical directions. Note also that the ratio of the characteristic values of the mean velocity to the turbulent velocity in the central part of the chamber varied from $1$ at $f=0$ to $0.5$ when the frequency $f$ of the grid oscillations $f > 2$ Hz. In particular, the characteristic values of the mean velocity at the frequencies of the grid oscillations $f = 0; 2.2; 16$ Hz are as following: $7.5$ cm/s; $3$ cm/s and $6$ cm/s, respectively.

\begin{figure}
\vspace*{1mm}
\centering
\includegraphics[width=9cm]{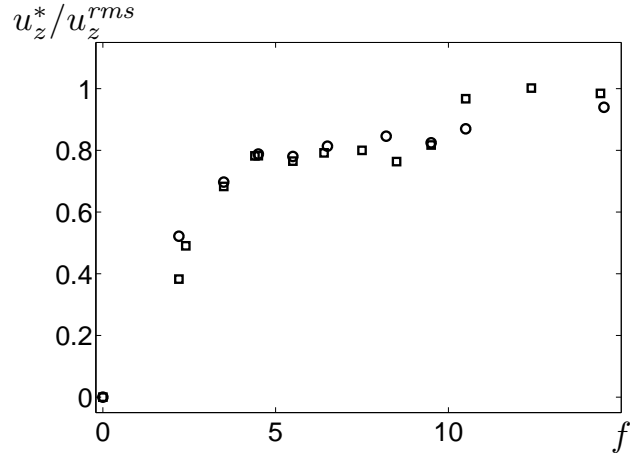}
\caption{\label{Fig10} Ratios $u^\ast_z/u_z^{rms}$ (squares) and $u^\ast_z/\tilde u_z$ (circles) versus the frequency $f$ of the grid oscillations. Here $u_z^{rms}$ is the r.m.s. of the vertical component of velocity fluctuations in the unstably stratified turbulent flow, $u^\ast_z$ is the r.m.s. of the vertical component of velocity fluctuations in the isothermal turbulence,  $\tilde u_z$ is the vertical component of the effective turbulent velocity, $\tilde u_z = [(u^\ast_z)^2 + 4 \ell_z  \, \beta \, \sqrt{\langle \theta^2 \rangle}]^{1/2}$, that takes into account the production of the turbulence by buoyancy. The velocity is measured in cm s$^{-1}$ and the frequency $f$ is measured in Hz.}
\end{figure}

In Fig.~\ref{Fig10} we show the frequency dependence of the ratio $u^\ast_z/u_z^{rms}$, where $u_z^{rms}$ is the r.m.s. of the vertical component of velocity fluctuations in the unstably stratified turbulent flow and $u^\ast_z$ is the r.m.s. of the vertical component of velocity fluctuations in the isothermal turbulence. In Fig.~\ref{Fig10} along with the ratios of the r.m.s. of the vertical component of velocity fluctuations, $u_z^\ast/u_z^{rms}$, in isothermal and unstably stratified turbulent flows, we also plot the ratio $u_z^\ast/\tilde u_z$, where $\tilde u_z$ is the vertical component of the effective turbulent velocity,
\begin{eqnarray}
\tilde u_z = [(u^\ast_z)^2 + 4 \ell_z \, \beta \, \sqrt{\langle \theta^2 \rangle}]^{1/2}  \;,
\label{FA3}
\end{eqnarray}
$\beta=g/T_\ast$  is the buoyancy parameter, $T_\ast$ is a reference value of the mean absolute temperature and ${\bf g}$ is the acceleration of gravity. This definition of the effective velocity is derived from the budget equation~(\ref{FA2}) for the turbulent kinetic energy in Sect.~III.
This effective velocity takes into account the production of the turbulence by buoyancy. Inspection of Fig.~\ref{Fig10} shows that the values of these ratios are very close. The latter implies that the measured turbulent velocity in the unstably stratified turbulent flow, $u_z^{rms}$, is of the order of $\tilde u_z$.

\begin{figure}
\vspace*{1mm}
\centering
\includegraphics[width=9cm]{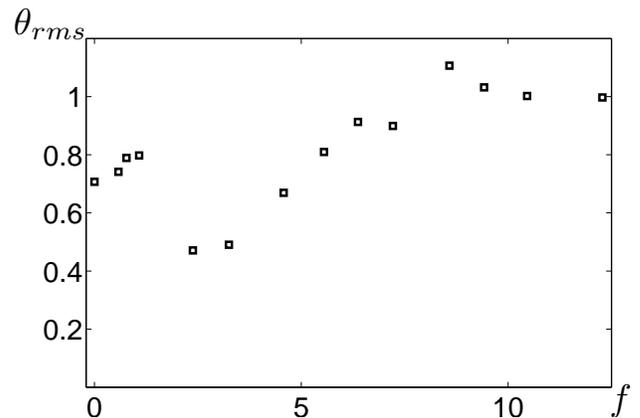}
\caption{\label{Fig11} The r.m.s. of temperature fluctuations $\theta_{rms}=\sqrt{\langle \theta^2 \rangle}$ versus the frequency $f$ of the grid oscillations for the unstably stratified turbulent flow.}
\end{figure}

The dependence of the r.m.s. of the temperature fluctuations $\sqrt{\langle \theta^2 \rangle}$ versus the frequency $f$ of the grid oscillations is shown in Fig.~\ref{Fig11}. This dependence is caused by the nontrivial behaviour of the mean temperature gradients along $x, y, z$ axes (see Figs.~\ref{Fig3}-\ref{Fig5}), where $\theta$ are fluctuations of fluid temperature. We have also analyzed the experimental results obtained at the smaller temperature difference, $\Delta T=25$ K, between the top and bottom walls (i.e., smaller Ra), and found only minor changes in the final results.
In the next section we analyze the obtained experimental results.

\section{Theoretical analysis and comparison with experimental results}

Let us first analyze the frequency dependence of the temperature fluctuations. To this end we consider the budget equation for $\langle \theta^2 \rangle$:
\begin{eqnarray}
{D \langle \theta^2 \rangle \over Dt} + {\rm div} \, {\bf \Phi}_\theta &=& - 2 ({\bf F} {\bf \cdot} \bec{\nabla}) T - 2 \varepsilon_\theta \;,
\label{A1}
\end{eqnarray}
(see, e.g., \cite{KF84,CCH02,ZEKR07,ZEKR08}), where $D / Dt = \partial / \partial t + {\bf U} {\bf \cdot} \bec{\nabla} $, $\; {\bf u}$ are the fluctuations of the fluid velocity, ${\bf U}$ is the mean velocity that describes the coherent structures, ${\bf \Phi}_\theta = \langle {\bf u} \,  \theta^2 \rangle$ is the third-order moment that determines the flux of $\langle \theta^2 \rangle$, $\, F_i = \langle u_i \theta \rangle$ is the turbulent heat flux, $\theta$ are the temperature fluctuations, $T$ is the mean temperature, $\varepsilon_\theta = D \langle (\bec{\nabla} \theta)^2\rangle$ is the dissipation rate of $\langle \theta^2 \rangle$, where $D$ is the coefficient of molecular temperature diffusion.

Let us use the Kolmogorov-Obukhov hypothesis which allows to estimate the dissipation rate $\varepsilon_\theta$ of $\langle \theta^2 \rangle$, as $\varepsilon_\theta \approx \langle \theta^2 \rangle / \tau_0$ (see, e.g., \cite{MY75,Mc90}), where $\tau_0=\ell/u_0$ is the characteristic turbulent time and $u_0$ is the characteristic turbulent velocity at the scale $\ell$. Indeed, $\varepsilon_\theta \equiv D \langle (\bec{\nabla} \theta)^2\rangle = D \, \langle \theta^2 \rangle \, \int_{k_0}^{k_b} k^2 \, \tilde E_\theta(k) \,d k \approx \langle \theta^2 \rangle / \tau_0$, where $\tilde E_\theta(k)=(q-1) \, k_0^{-1} \, (k/k_0)^{-q}$ is the  spectrum function of the temperature fluctuations, $q$ is the exponent of the spectrum of the temperature fluctuations (e.g., $q=5/3$ for the  Corrsin-Obukhov spectrum), $k_0 = \ell^{-1}$, $\, k_b = \ell_b^{-1}$, $\ell_b=\ell / {\rm Pe}^{1/(3-q)}$ and ${\rm Pe} = u_{0} \, \ell / D$ is the Peclet number. The latter estimate implies that the main contribution to the dissipation rate $\varepsilon_\theta$ arises from very small molecular temperature diffusion scales $\ell_b$.
The Kolmogorov-Obukhov hypothesis has been widely used in atmospheric turbulence in a number of studies including investigations of atmospheric stably stratified and convective boundary layers.
In a steady-state Eq.~(\ref{A1}) yields:
\begin{eqnarray}
\langle \theta^2 \rangle &\approx& - 2 \tau_0 \, ({\bf F} {\bf \cdot} \bec{\nabla}) T
\nonumber\\
&\approx& 2 \big[(\ell_x \nabla_x T)^2 + (\ell_y \nabla_y T)^2 + (\ell_z \nabla_z T)^2\big] \, .
\label{A2}
\end{eqnarray}
In deriving Eq.~(\ref{A2}) we take into account that the components of the turbulent heat flux and turbulent temperature diffusion coefficients in an anisotropic turbulence are given by $F_x = - D^T_x \, \nabla_x T$ and $D^T_x = C_x \, \ell_x \, \sqrt{\langle u_x^2 \rangle}$, and similarly for other components $F_y, F_z$ of the turbulent heat flux and the turbulent temperature diffusion coefficients $D^T_y, D^T_z$ with the replacement in the above formulas $x$ by $y$ or by $z$. We also neglected a small term ${\rm div} \, {\bf \Phi}_\theta$ in Eq.~(\ref{A1}) for nearly homogeneous turbulent convection. When horizontal gradients of the mean temperature are much smaller than the vertical gradients, $|\nabla_{x,y} T| \ll |\nabla_z T|$, Eq.~(\ref{A2}) yields the following non-dimensional ratio
\begin{eqnarray}
{\ell_z \, |\nabla_z T| \over \sqrt{\langle \theta^2 \rangle}} = {\rm const} \; .
\label{A10}
\end{eqnarray}
Kraichnan (1968) derived Eq.~(\ref{A10}) for a passive scalar regime using the Lagrangian-history direct-interaction approximation for small Peclet numbers \cite{KR68}. Equation~(\ref{A10}) is in agreement with the corresponding equation derived by means of the path integral approach in \cite{EKR95} for the passive scalar advected by the delta-correlated in time Gaussian smooth velocity field. Equation~(\ref{A10}) can be also derived using spectrum of temperature fluctuations found by dimensional arguments in \cite{W58} and by the renormalization procedure in \cite{EKRS96}. In Appendix we also derive Eqs.~(\ref{A2}) and~(\ref{A10}) using other procedures, namely a quasi-linear approach (or a second order correlation approximation) that is valid for small Peclet numbers and the spectral $\tau$ approximation that is valid for large Peclet numbers.

Now let us discuss the generalization of Eq.~(\ref{A10}). To this end we use the budget equation for the turbulent heat flux $F_i = \langle u_i \theta \rangle$:
\begin{eqnarray}
{D F_i \over Dt} + {\nabla}_j \, {\bf \Phi}_{ij}^{(F)} &=& \beta_i \, \langle \theta^2 \rangle + {1\over \rho} \, \langle \theta \, \nabla_i p \rangle - \langle u_i u_j \rangle \, {\nabla}_j \, T
\nonumber\\
&&- ({\bf F} {\bf \cdot} \bec{\nabla}) U_i - \varepsilon_i^{(F)} \;,
\label{A11}
\end{eqnarray}
where $\beta_i = \beta \, e_z $, $\, p$ are the pressure fluctuations, $\rho$ is the fluid density, ${\bf U}$ is the mean velocity, $\varepsilon_i^{(F)}$ is the dissipation rate of the turbulent heat flux and ${\bf \Phi}_{ij}^{(F)} = \langle u_i u_j \theta\rangle + \delta_{ij} \rho^{-1} \, \langle \theta \, p \rangle / 2$ is the term that includes the third-order moments. According to the estimate made in \cite{ZEKR07}, $\beta_i \, \langle \theta^2 \rangle + \rho^{-1} \, \langle \theta \, \nabla_i p \rangle \approx 2 \beta_i \, \langle \theta^2 \rangle$. In a steady-state case Eqs.~(\ref{A1}) and~(\ref{A11}) yield:
\begin{eqnarray}
{\ell_\ast \, \nabla_\ast T \over \sqrt{\langle \theta^2 \rangle}} = {1 \over 2 C_\theta C_F} = {\rm const} \;,
\label{A12}
\end{eqnarray}
where
\begin{eqnarray}
[\ell_\ast \, \nabla_\ast T]^2 &=& \big[(\ell_x \nabla_x T)^2 + (\ell_y \nabla_y T)^2 + (\ell_z \nabla_z T)^2\big]
\nonumber\\
&& \times \big[1 + 4 C_\theta C_F \beta \tau_0^2 (\nabla_z T)\big]^{-1} \;,
\label{A14}
\end{eqnarray}
and $C_F$ is an empirical constant. Here we take into account that the dissipation rate of the turbulent heat flux is $\varepsilon_i^{(F)} = F_i / C_F \tau_0$ and $\tau_x \approx \tau_y \approx \tau_z = \tau_0$. In Eqs.~(\ref{A12}) and~(\ref{A14}) we also have neglected the terms $\sim O[\ell^3/L_T^3; \ell^3/(L_T^2 L_U)]$, where $L_T$ and $L_U$ are the characteristic spatial scales of the mean temperature and velocity field variations. Equation~(\ref{A12}) is valid in a general case including the passive scalar regime. In the case of the passive scalar regime whereby there is only the vertical gradient of the mean temperature, Eq.~(\ref{A12}) coincides with Eq.~(\ref{A10}).

\begin{figure}
\vspace*{1mm}
\centering
\includegraphics[width=9cm]{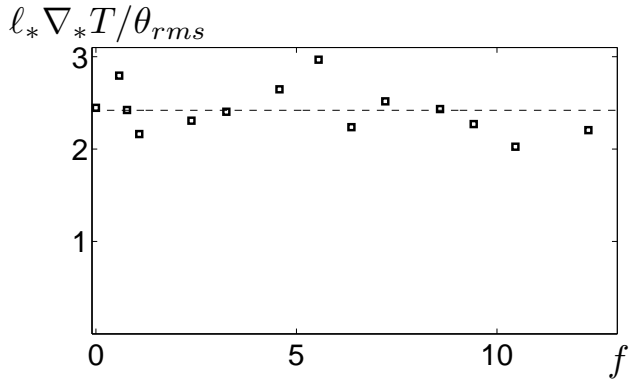}
\caption{\label{Fig12} The non-dimensional ratio $\ell_\ast \, \nabla_\ast T / \theta_{rms}$ versus the frequency $f$ of the grid oscillations for the unstably stratified turbulent flow, where $\theta_{rms}=\sqrt{\langle \theta^2 \rangle}$.}
\end{figure}

In Fig.~\ref{Fig12} we plot the non-dimensional ratio $\ell_\ast \, \nabla_\ast T / \sqrt{\langle \theta^2 \rangle}$ versus the frequency $f$ of the grid oscillations for the unstably stratified turbulent flow obtained in our experiments, where we assumed that $\ell_x = \ell_y$ and $C_\theta C_F = 1/12$. Inspection of Fig.~\ref{Fig12} shows that this non-dimensional ratio varies slightly even in the range of parameters whereby the behaviour of the temperature field is different from that of the passive scalar. Small deviations of the experimental results from the theoretical predictions [see Eq.~(\ref{A12})] may be caused by a non-zero term, ${\rm div} \, {\bf \Phi}_\theta$ and deviations from the steady-state due to the long-term oscillations of the mean temperature.

Now let us derive Eq.~(\ref{FA3}) for the measured turbulent velocity in the unstably stratified turbulent flow. We use the budget equation for the turbulent kinetic energy $E_k=\langle {\bf u}^2 \rangle/2$:
\begin{eqnarray}
{DE_k \over Dt} + {\rm div} \, {\bf \Phi}_{k} &=& -\langle u_i \, u_j \rangle \, \nabla_j U_i + \beta \, F_z + \langle {\bf u} {\bf \cdot} {\bf f}_f \rangle - \varepsilon_k ,
\nonumber\\
\label{FA2}
\end{eqnarray}
where ${\bf \Phi}_{k} = \rho^{-1} \langle {\bf u} \, p\rangle + (1/2) \langle {\bf u} \, u^2 \rangle$ is the term that includes the third-order moments, $\langle {\bf u} {\bf \cdot} {\bf f}_f \rangle$ is the production of turbulence caused by the grid oscillations and $\varepsilon_k$ is the dissipation rate of the turbulent kinetic energy.  In the steady-state Eq.~(\ref{FA2}) yields $\langle{\bf u}^2 \rangle = 2 \tau_0 \, [\nu_{_{T}} \, S^2 + \langle {\bf u} {\bf \cdot} {\bf f}_f \rangle + \beta \, F_z]$, where $\nu_{_{T}}$ is the turbulent viscosity. Let us introduce the characteristic velocity for isothermal turbulence $(u^\ast)^2 = 2 \tau_0 \, [\nu_{_{T}} \, S_\ast^2 + \langle {\bf u} {\bf \cdot} {\bf f}_f \rangle]$, where $S_\ast$ is the large-scale shear that can appear in isothermal oscillating grid turbulence. Note that for large frequencies of the grid oscillations the large-scale circulations caused by turbulent convection disappear and $S \to S_\ast$. Since the vertical turbulent heat flux $F_z = - D_T \nabla_z T$ and for large frequencies of the grid oscillations $|\nabla_z T| \sim \sqrt{\langle \theta^2 \rangle} / \ell_z$ [see Eq.~(\ref{A10})], we arrive at the following equation:
\begin{eqnarray}
u_z^{rms} \sim [(u^\ast_z)^2 + 4 \ell_z \, \beta \, \sqrt{\langle \theta^2 \rangle}]^{1/2} \;,
\label{FFA}
\end{eqnarray}
where we take into account that the r.m.s. of the horizontal component of velocity fluctuations in the unstably stratified turbulent flow $\sqrt{\langle {\bf u}^2_h \rangle}  \simeq u_h^\ast$ and $u_h^\ast = [(u^\ast)^2-(u_z^\ast)^2]^{1/2}$. This is the reason that the frequency dependencies of the ratios $u^\ast_z/u_z^{rms}$ and $u^\ast_z/\tilde u_z$ obtained in our experiments (see Fig.~\ref{Fig10}), are close.

Using Eq.~(\ref{A1}) we obtain the evolutionary equation for the turbulent potential energy $E_p= (\beta^2 /2 N^2) \, \langle \theta^2 \rangle$:
\begin{eqnarray}
{D E_p \over Dt} + {\rm div} \, {\bf \Phi}_p &=& P_p - \beta F_z - \varepsilon_p ,
\label{PA1}
\end{eqnarray}
(see, e.g., \cite{ZEKR07,ZEKR08}), where $N^2= \beta \, \nabla_z T$, $\, {\bf \Phi}_p= (\beta^2 /2 N^2) \, {\bf \Phi}_\theta$, $\, P_p= - (\beta^2 /N^2) \, ({\bf F}_h {\bf \cdot} \bec{\nabla}) T$ is the source (or sink) of the turbulent potential energy caused by the horizontal turbulent heat flux ${\bf F}_h=\langle {\bf u}_h \, \theta \rangle$, $\, {\bf u}_h$ is the horizontal component of the velocity fluctuations and $\varepsilon_p= (\beta^2 /2 N^2) \, \varepsilon_\theta$. The turbulent potential energy is analogous to the available potential energy introduced by Lorenz \cite{L55}: the available potential energy and the turbulent potential energy are proportional to the squared temperature fluctuations. The principal difference between these two quantities is that the available potential energy is an integral property of the entire flow-domain (e.g., of the atmosphere as a whole), whereas the turbulent potential energy is a local quantity, i.e., it is determined in each point of turbulent flow.
The buoyancy term, $\beta \, F_z$, appears in Eqs.~(\ref{FA2}) and~(\ref{PA1}) with opposite signs and describes the energy exchange between the turbulent kinetic energy and the turbulent potential energy. These two terms cancel in the budget equation for the total turbulent energy, $E=E_k+E_p$:
\begin{eqnarray}
{D E \over Dt} + {\rm div} \, {\bf \Phi} &=& P_p -\langle u_i \, u_j \rangle \, \nabla_j U_i + \langle {\bf u} {\bf \cdot} {\bf f}_f \rangle - \varepsilon ,
\label{PA2}
\end{eqnarray}
where ${\bf \Phi}={\bf \Phi}_k + {\bf \Phi}_p$ and $\varepsilon= \varepsilon_k + \varepsilon_p$. The concept of the total turbulent energy is very useful in analysis of stratified turbulent flows. In particular, it allows to explain the physical mechanism for existence of turbulence for arbitrary values of the Richardson number, and abolish the paradigm of the critical Richardson number in the stably stratified atmospheric turbulence (see \cite{ZEKR07,ZEKR08}).

\begin{figure}
\vspace*{1mm}
\centering
\includegraphics[width=9cm]{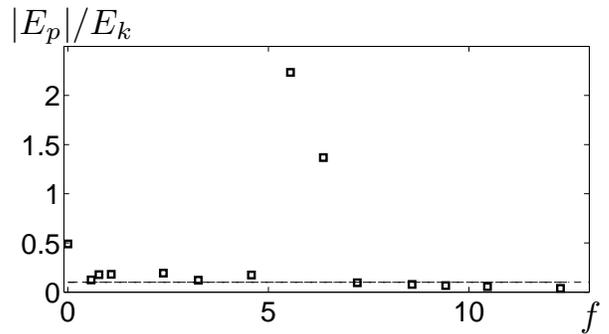}
\caption{\label{Fig13} The non-dimensional ratio $|E_p|/E_k$ of the potential to kinetic energies versus the frequency $f$ of the grid oscillations for the unstably stratified turbulent flow. The horizontal line corresponds to $|E_p|/E_k=0.1$.}
\end{figure}

\begin{figure}
\vspace*{1mm}
\centering
\includegraphics[width=9cm]{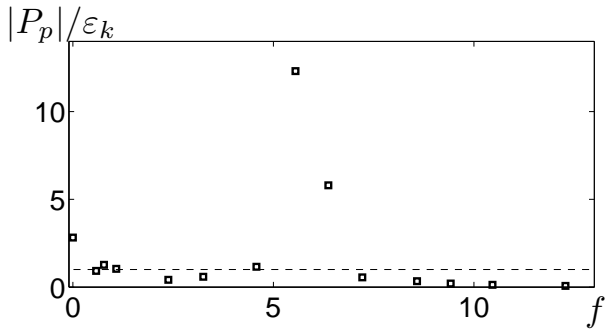}
\caption{\label{Fig14} The non-dimensional ratio $|P_p|/\varepsilon_k$ of the source (or sink) of the turbulent potential energy caused by the horizontal turbulent heat flux, to the dissipation rate of the turbulent kinetic energy versus the frequency $f$ of the grid oscillations for the unstably stratified turbulent flow. The horizontal line corresponds to $|P_p|/\varepsilon_k=1$.}
\end{figure}

For very small horizontal mean temperature gradients, the source term $P_p$ is much smaller than other terms in the right hand side of Eq.~(\ref{PA2}). In Figs.~\ref{Fig13} and~\ref{Fig14} we show
the non-dimensional ratios $|E_p|/E_k$ and $|P_p|/\varepsilon_k$ versus the frequency $f$ of the grid oscillations for the unstably stratified turbulent flow. Inspection of Figs.~\ref{Fig13} and~\ref{Fig14} shows that the temperature fluctuations affect the turbulent kinetic energy only in the vicinities of $f=0$ and $f=5.2$ Hz. Note that in the vicinity of $f=5.2$ Hz, the sign of $\nabla_z T$ changes from positive to negative (see Fig.~\ref{Fig3}).

In view of the above analysis, we consider the passive-like scalar behaviour of  the temperature field in the kinematic sense. In particular, if the temperature fluctuations $\langle \theta^2 \rangle$ do not affect the turbulent kinetic energy, the temperature field can be considered as a passive scalar. In this case the evolution of the temperature field in a given turbulent velocity field is a kinematic problem, whereby there is no dynamic coupling between the temperature fluctuations, $\langle \theta^2 \rangle$, and the turbulent kinetic energy in spite of a non-zero turbulent heat flux, i.e., non-zero correlations of the velocity and temperature fluctuations. When the effect of the temperature fluctuations on the turbulent kinetic energy cannot be neglected, the temperature is considered as an active field.
This understanding of the passive or active behavior of the temperature field is different from that based on the scaling behaviour of the temperature structure function (see \cite{LX10}).

Note that the turbulent convection studied in our experiments is non-Boussinesq, i.e., div $\, {\bf u} \not = 0$. In the theoretical analysis we use the anelastic approximation, div $\, (\rho \, {\bf u}) = 0$. It should be noted that the deviation from the Boussinesq approximation in our experiments is less than 10 \% (i.e., spatial variations of the fluid density are $\delta \rho / \rho < 0.1$).

\section{Conclusions}

In the present study we have investigated transition phenomena caused by the external forcing from Rayleigh-B\'{e}nard convection with LSC to the limiting regime of unstably stratified turbulent flow without LSC, whereby the temperature field behaves like a passive scalar. The external forcing of turbulence is produced by two oscillating grids. When the frequency of the grid oscillations is larger than $2$ Hz, the large-scale circulation (LSC) in the $yz$-plane in turbulent convection is destroyed. On the other hand, a complicated mean flow of the spiral form in the $xz$-plane with a non-zero mean vorticity in the $y$ direction exists when the frequencies of the grid oscillations $2<f<7$ Hz. The destruction of the LSC is accompanied by a strong change of the mean temperature distribution. In the central part of the flow the vertical mean temperature gradient changes its direction depending on the frequency of the grid oscillations.
We have shown that for all regimes of the unstably stratified turbulent flow the ratio $\big[(\ell_x \nabla_x T)^2 + (\ell_y \nabla_y T)^2 + (\ell_z \nabla_z T)^2\big] / \langle \theta^2 \rangle$ varies slightly even in the range of parameters whereby the behaviour of the temperature field is different from that of the passive scalar.
The experimental results obtained in this study are in an agreement with the theoretical predictions based on the budget equations for turbulent kinetic energy, turbulent potential energy and turbulent heat flux.

We have also found that in the unstably stratified turbulent flow the vertical component of the turbulent velocity field is determined by Eq.~(\ref{FFA}). The latter equation is derived from the budget  equation for the turbulent kinetic energy. This implies that the buoyancy heat flux contributes to the production of the turbulent kinetic energy in turbulent convection in the absence of the LSC. However, the turbulent kinetic energy in large Rayleigh number turbulent convection with coherent structures (LSC) is produced by shear, rather than by buoyancy (see \cite{BEKR09}).

The long-term nonlinear oscillations of the mean temperature have been detected for all regimes, from the turbulent convection with LSC to the passive scalar regime, at all frequencies of the grid oscillations.

\appendix

\section{Temperature fluctuations}

Evolution of  the instantaneous temperature field $T_{\rm tot}(t, {\bf r})$ in a turbulent flow is determined by the following equation:
\begin{eqnarray}
{\partial T_{\rm tot} \over \partial t} + (\hat{\bf v} {\bf \cdot} \bec\nabla) \, T_{\rm tot} + (\gamma - 1) \, ({\bf  \nabla} {\bf \cdot} \hat{\bf v}) \, T_{\rm tot} = D \,\bec{\nabla}^2 T_{\rm tot} \;,
\nonumber\\
\label{BBB1}
\end{eqnarray}
where $T_{\rm tot}(t, {\bf r})= T +\theta$, $\, D$ is the coefficient of molecular temperature diffusion, $\gamma=c_{\rm p}/c_{\rm v}$ is the ratio of specific heats,  $\hat{\bf v}$ is the velocity field that satisfies to continuity equation in anelastic approximation for a low-Mach-number flow:
\begin{eqnarray}
{\bf  \nabla} {\bf \cdot} (\rho \, \hat{\bf v}) =0 \; .
\label{AB1}
\end{eqnarray}
Combining Eq.~(\ref{BBB1}) and~(\ref{AB1}) we obtain the following equation:
\begin{eqnarray}
{\partial T_{\rm tot} \over \partial t} + ({\bf v} {\bf \cdot} \bec\nabla) \, T_{\rm tot} = D \,\bec{\nabla}^2 T_{\rm tot} \;,
\nonumber\\
\label{BB1}
\end{eqnarray}
where ${\bf v}= \gamma \hat{\bf v}$ and we take into account that for a low-Mach-number flow without  imposed external pressure gradient $\bec\nabla \rho/\rho \approx - \bec\nabla T_{\rm tot}/T_{\rm tot}$.

Averaging Eq.~(\ref{BB1}) over ensemble of turbulent velocity field we obtain the equation for the evolution of the mean temperature field $T(t, {\bf r})$:
\begin{eqnarray}
{\partial T \over \partial t} + ({\bf U} {\bf \cdot} \bec\nabla) \, T + \bec\nabla {\bf \cdot} \, {\bf F} = D \,\bec{\nabla}^2 T \;,
\label{BB4}
\end{eqnarray}
where ${\bf U} = \gamma \hat{\bf U}$, $\, \hat{\bf U}$  is the mean fluid velocity, ${\bf F} = \langle {\bf u} \, \theta \rangle$ is the heat flux, ${\bf u} = \gamma \hat{\bf u}$ and $\hat{\bf u}$ are the velocity fluctuations.

\subsection{Temperature fluctuations for small Peclet numbers}

In order to study temperature fluctuations for small Peclet numbers  we use a quasi-linear approach or a second order correlation approximation (see, e.g., \cite{M78}). Subtracting Eq.~(\ref{BB4}) from Eq.~(\ref{BB1}) yields equation for the temperature fluctuations:
\begin{eqnarray}
{\partial \theta \over \partial t} + Q - D \bec{\nabla}^2 \theta  = I \,,
\label{C1}
\end{eqnarray}
where $I = - ({\bf u} {\bf \cdot} \bec{\nabla}) T$ is the source term and $Q = \bec\nabla {\bf \cdot} \, [{\bf u} \theta - \langle {\bf u} \theta \rangle] + \theta ({\bf u} {\bf \cdot} \bec\nabla) \rho/ \rho$  is the nonlinear term. In the expression for $Q$ we neglected the term $\langle \theta {\bf u} \rangle {\bf \cdot} \bec\nabla \rho/ \rho$ which is quadratic in large-scale spatial derivative.
Let us neglect the nonlinear term but keep the molecular diffusion term in Eq.~(\ref{C1}). For this reason this approach is called a quasi-linear or perturbation approach. This approximation for a given velocity field is valid only for small Peclet numbers (${\rm Pe} \ll 1$), where ${\rm Pe} = u_{0} \, \ell / D$. Let us rewrite Eq.~(\ref{C1}) in a Fourier space. Then the solution of Eq.~(\ref{C1}) is given by $\theta(\omega, {\bf k}) = G_D(\omega, {\bf k}) I(\omega, {\bf k})$, where $G_D(\omega, {\bf k}) = (D k^2 + i \omega)^{-1}$.

Let us apply a standard two-scale approach, whereby the non-instantaneous two-point second-order correlation function
is written as follows:
\begin{eqnarray}
&& \langle u_i(t_1, {\bf x}) \, \theta(t_2, {\bf  y}) \rangle = \int \langle u_i (\omega_1, {\bf k}_1) \theta(\omega_2, {\bf k}_2) \rangle \exp[i({\bf  k}_1 {\bf \cdot} {\bf x}
\nonumber\\
&& \quad \quad + {\bf k}_2 {\bf \cdot} {\bf y}) + i(\omega_1 t_1 + \omega_2 t_2)] \,d\omega_1 \, d\omega_2 \,d{\bf k}_1 \, d{\bf k}_2
\nonumber\\
&& \quad \quad = \int F_i(\omega, {\bf k})  \exp[i {\bf k} {\bf \cdot} {\bf r} + i\omega \, \tilde \tau] \,d\omega \,d {\bf k} \,,
\label{C2}
\end{eqnarray}
where we use large scale variables: ${\bf R} = ({\bf x} + {\bf y}) / 2$, $\, {\bf K} = {\bf k}_1 + {\bf k}_2$, $\, t = (t_1 + t_2) / 2$, $\, \Omega = \omega_1 + \omega_2$, and small scale  variables: ${\bf r} = {\bf x} - {\bf y}$, $\, {\bf k} = ({\bf k}_1 - {\bf k}_2) / 2$, $\, \tilde \tau = t_1 - t_2$, $\, \omega = (\omega_1 - \omega_2) / 2$,
\begin{eqnarray}
&& F_i(\omega, {\bf k}) = \int \langle u_i(\omega_1, {\bf k}_1) \, \theta(\omega_2, {\bf k}_2) \rangle \exp[i \Omega t
\nonumber\\
&& \quad \quad\quad\quad + i {\bf K} {\bf \cdot} {\bf R}] \,d \Omega \,d {\bf  K} \,,
\label{C3}
\end{eqnarray}
and $\omega_1 = \omega + \Omega / 2$, $\, \omega_2 = - \omega + \Omega / 2$, ${\bf k}_1 = {\bf k} + {\bf  K} / 2$, $\, {\bf k}_2 = - {\bf k} + {\bf  K} / 2$ (see, e.g., \cite{RS75}). We assume here that there exists a separation of scales,
i.e., the maximum scale of random motions $\ell$ is much
smaller than the characteristic scales of inhomogeneities of the
mean temperature and fluid density.

The turbulent heat flux $F_i$ and the function $E_\theta=\langle \theta^2 \rangle/2$ are given by the following relations:
\begin{eqnarray}
& & F_i = \int \langle u_i(\omega, {\bf k}) \, I(- \omega, - {\bf k}) \rangle \, G_D^\ast \,d \omega \,d {\bf  k} \,,
\label{C5}\\
& & E_\theta = {1 \over 2} \int \langle u_i(\omega, {\bf k}) \, u_j(- \omega, - {\bf k}) \rangle \, G_D \, G_D^\ast \,d \omega \,d {\bf  k}
\nonumber\\
& & \quad \quad \quad \times (\nabla_i T) \, (\nabla_j T) \, .
\label{C8}
\end{eqnarray}

Hereafter we use the simple model for the second moments of a random velocity field, $\langle u_i(\omega, {\bf k}) \, u_j(-\omega, -{\bf k}) \rangle$, in a Fourier space:
\begin{eqnarray}
&& \langle u_i(\omega, {\bf k}) \, u_j(-\omega, -{\bf k}) \rangle = {\gamma^2 \,u_0^2 \, E(k) \over 8 \pi k^2} \Big[\delta_{ij} - {k_i \, k_j \over k^2} \Big] \, \delta(\omega) \,,
\nonumber\\
\label{C4}
\end{eqnarray}
where $\delta(\omega)$ is the Dirac's delta-function, $\, \delta_{ij}$ is the Kronecker tensor, $P_{ij}(k)=\delta_{ij} - k_i k_j /  k^2$, the energy spectrum function is $E(k) = k_0^{-1} \, (q-1) \, (k / k_{0})^{-q}$, the exponent $1<q<3$, the wave number $k_{0} = 1 / \ell$, the length $\ell$ is the maximum scale of random motions and $u_0$ is the characteristic velocity in the maximum scale of random motions.
Note that contributions due to the density stratification in the random velocity field, $\langle u_i(\omega, {\bf k}) \, u_j(-\omega, -{\bf k}) \rangle$ to the turbulent heat flux $F_i$ and the temperature fluctuations $\langle \theta^2 \rangle$ vanish.

After integration in $\omega$-space and in ${\bf k}$-space, Eqs.~(\ref{C5}) and~(\ref{C8}) yield formulae for the turbulent heat flux $F_i$ and the temperature fluctuations $\langle \theta^2 \rangle$
in an isotropic background turbulence:
\begin{eqnarray}
\langle u_i \, \theta \rangle &=& - D_T \nabla_i T \;, \quad D_T = C_D \, u_{0} \, \ell \,,
\label{C6}\\
\langle \theta^2 \rangle &=& - C_\ast \, \ell^2 \, (\bec{\nabla} T)^2 \,,
\label{C7}\\
C_D &=& {(q-1) \, \gamma^2 \over 3(q+1)} \, {\rm Pe} \,, \;\; C_\ast = {(q-1) \, \gamma^2 \over 3(q+3)} \, {\rm Pe}^2\,,
\nonumber
\end{eqnarray}
where $D_T$ is the turbulent temperature diffusion coefficient.
Equations~(\ref{C6}) and~(\ref{C7}) also can be obtained using spectrum of temperature fluctuations for small Peclet numbers found by means of dimensional arguments in \cite{BH59} and by the Lagrangian-history direct-interaction approximation applied in \cite{KR68}.

\subsection{Temperature fluctuations for large Peclet numbers}

In this subsection we derive formulae for the turbulent heat flux $F_i$ and the temperature fluctuation function $E_\theta$ using the $\tau$ approach that is valid for large Peclet and Reynolds numbers. Using Eq.~(\ref{C1}) written in a Fourier space we derive equation for the instantaneous two-point second-order correlation functions $F_i(t, {\bf k}) = \langle u_i(t, {\bf k}) \, \theta(t, -{\bf k}) \rangle$ and $E_\theta(t, {\bf k}) = (1/2) \, \langle \theta(t, {\bf k}) \, \theta(t, -{\bf k}) \rangle$:
\begin{eqnarray}
{dF_i \over dt} &=& \langle u_i(t, {\bf k}) \, I(t, -{\bf k}) \rangle + \hat{\cal M} F_i^{(III)}({\bf k}) \,,
\label{BD1}\\
{dE_\theta \over dt} &=& I_\theta(t, {\bf k}) + \hat{\cal M} E_\theta^{(III)}({\bf k}) \,,
\label{BD2}
\end{eqnarray}
where $\hat{\cal M} F_i^{(III)}({\bf k}) = - [\langle u_i \, Q \rangle + \langle (\partial u_i / \partial t) \, \theta \rangle - D \langle u_i \, \bec{\nabla}^2 \theta \rangle]_{\bf k}$ and $\hat{\cal M} E_\theta^{(III)}({\bf k}) = - (1/2) \, [\langle \theta \, Q \rangle - D \langle \theta \, \bec{\nabla}^2 \theta \rangle]_{\bf k}$ are the third-order moment terms appearing due to the nonlinear terms which include also molecular diffusion term, and
\begin{eqnarray*}
I_\theta(t, {\bf k}) = {1 \over 2} \left[\langle I(t, {\bf k}) \, \theta(t, -{\bf k}) \rangle + \langle \theta(t, {\bf k}) \, I(t, -{\bf k}) \rangle \right] \, .
\end{eqnarray*}

The equation for the second moment includes the first-order spatial
differential operators $\hat{\cal M}$  applied to the third-order
moments $F^{(III)}$. A problem arises how to close the system, i.e.,
how to express the third-order terms $\hat{\cal M}
F^{(III)}$ through the lower moments $F^{(II)}$
(see, e.g., \cite{O70,MY75,Mc90}). We use the spectral $\tau$ approximation which postulates that the deviations of the third-moment terms, $\hat{\cal M} F^{(III)}({\bf k})$, from the contributions to these terms afforded by the background turbulence, $\hat{\cal M} F^{(III,0)}({\bf k})$, can be expressed through the similar deviations of the second moments, $F^{(II)}({\bf k}) - F^{(II,0)}({\bf k})$:
\begin{eqnarray}
&& \hat{\cal M} F^{(III)}({\bf k}) - \hat{\cal M} F^{(III,0)}({\bf
k})
\nonumber\\
&& \quad \quad\quad = - {1 \over \tau_r(k)} \, \Big[F^{(II)}({\bf k}) - F^{(II,0)}({\bf k})\Big] \,,
\label{D2}
\end{eqnarray}
(see, e.g., \cite{O70,PFL76,RK07}), where $\tau_r(k)$ is the scale-dependent relaxation time, which can be identified with the correlation time $\tau(k)$ of the turbulent velocity field for large Reynolds and Peclet numbers. The functions with the superscript $(0)$ correspond to the background turbulence with a zero gradient of the mean temperature. Validation of the $\tau$ approximation for different situations has been performed in numerous numerical simulations and analytical studies (see, e.g., review \cite{BS05}; and also discussion in \cite{RK07}, Sec. 6).

Note that the contributions of the terms with the superscript $(0)$ vanish because when the gradient of the mean temperature is zero, the turbulent heat flux and the temperature fluctuations vanish. Consequently, Eq.~(\ref{D2}) for $\hat{\cal M} F_i^{(III)}({\bf k})$ reduces to $\hat{\cal M} F_i^{(III)}({\bf k}) = - F_i({\bf k}) / \tau(k)$ and for $\hat{\cal M} E_\theta^{(III)}({\bf k}) = - E_\theta({\bf k}) / \tau(k)$.
We also assume that the characteristic time of variation of the second moments $F_i({\bf k})$ and $E_\theta({\bf k})$ are substantially larger than the correlation time $\tau(k)$ for all turbulence scales. Therefore, in a steady-state Eqs.~(\ref{BD1}) and~(\ref{BD2}) yield the following formulae for the turbulent heat flux $F_i$ and the function $E_\theta$:
\begin{eqnarray}
F_i &=& \int \tau(k) \, \langle u_i(t, {\bf k}) \, I(t, -{\bf k}) \rangle \, d{\bf k} \,,
\label{BD3}\\
E_\theta &=& \int \tau(k) \, I_\theta(t, {\bf k}) \, d{\bf k} \, .
\label{BD4}
\end{eqnarray}
Now we use the following simple model for the second moments of turbulent velocity field, $\langle u_i({\bf k}) \, u_j(-{\bf k}) \rangle$ in ${\bf k}$ space:
\begin{eqnarray}
\langle u_i({\bf k}) \, u_j(-{\bf k}) \rangle = {\gamma^2 \, u_0^2 \, E(k) \over 8 \pi k^2} \Big[\delta_{ij} - {k_i \, k_j \over k^2} \Big]  \, .
\label{D3}
\end{eqnarray}
Contributions due to the fluid density stratification in the random velocity field, $\langle u_i({\bf k}) \, u_j(-{\bf k}) \rangle$ to the turbulent heat flux $F_i$ and the temperature fluctuations $\langle \theta^2 \rangle$ vanish.
After integration in ${\bf k}$-space Eqs.~(\ref{BD3}) and~(\ref{BD4}) we arrive at equations for the turbulent heat flux, $F_i$, and the temperature fluctuations, $E_\theta$ in an isotropic background turbulence:
\begin{eqnarray}
\langle u_i \, \theta \rangle &=& - D_T \nabla_i T \;, \quad D_T = C_D \, u_{0} \, \ell \,,
\label{LC6}\\
\langle \theta^2 \rangle &=& - (8/9) \, \ell^2 \, (\bec{\nabla} T)^2   \, .
\label{LC7}
\end{eqnarray}
where $C_D=\gamma^2 / 3$.
In the derivation we used the following expression for the turbulent correlation time, $\tau(k) = 2 \, \tau_0 \, (k / k_{0})^{1-q}$, where $\tau_0 = \ell / u_{0}$ is the characteristic turbulent time. Therefore, the formulae for the turbulent heat flux $F_i$ and the function $E_\theta$ are similar for small and large Peclet numbers, while the coefficients are different in these two cases. Notably, the theoretical predictions for large Peclet number (which correspond to the laboratory experiments) are in agrement with experimental results and with the estimate based on the balance equation for $E_\theta$ [see Eq.~(\ref{A2})]. Equations~(\ref{LC6}) and~(\ref{LC7}) are in agreement with the corresponding equations derived by means of the path integral approach in \cite{EKR95} for the passive scalar advected by the delta-correlated in time Gaussian smooth velocity field. Equations~(\ref{LC6}) and~(\ref{LC7}) also can be obtained using spectrum of temperature fluctuations found by means of dimensional arguments in \cite{W58} and by the renormalization procedure for large Peclet numbers performed in \cite{EKRS96}.

Note that in our experiments with large frequencies of the grid oscillations, i.e., in the limiting regime of the unstably stratified turbulent flow without the LSC, whereby the temperature field behaves like a passive scalar, the constant $C_D$ is nearly independent of the molecular heat transport coefficients for large Peclet numbers (which correspond to the conditions of our experiments). The constant $C_D$ depends on the molecular heat transport coefficients only for small Peclet numbers.

\begin{acknowledgements}
We thank A.~Krein for his assistance in construction of the
experimental set-up and Y. Gluzman for his assistance in processing of the experimental results. This research was supported in part by the Israel Science Foundation governed by the Israeli Academy of Sciences (Grant 259/07).
\end{acknowledgements}

\end{document}